\documentclass[9pt,twocolumn]{article}
\usepackage{arxiv}
\usepackage{authblk} 
\usepackage[utf8]{inputenc}
\usepackage{graphicx}
\usepackage{amssymb}

\usepackage{url,hyperref,microtype,subcaption}
\usepackage[onehalfspacing]{setspace}
\usepackage{lipsum}
\usepackage{booktabs}		
\usepackage{soul}
\usepackage{multirow}		
\usepackage{amsmath}  

\DeclareMathOperator{\arctantwo}{arctan2}

\title{Impact of Variable Speed on Collective Movement of Animal Groups}
\author[1,2]{Pascal P. Klamser}
\author[1,3]{Luis G\'omez-Nava}
\author[3,4]{Tim Landgraf}
\author[5]{Jolle W. Jolles}
\author[3,6,7]{David Bierbach}
\author[1,2,3]{Pawel Romanczuk}

\affil[1]{Institute for Theoretical Biology, Department of Biology, Humboldt Universit\"at zu Berlin, Berlin, Germany.}
\affil[2]{Bernstein Center for Computational Neuroscience, 10115 Berlin, Germany.}
\affil[3]{Cluster of Excellence, Science of Intelligence, Technische Universität Berlin, Berlin, Germany.}
\affil[4]{Department of Mathematics and Computer Science, Freie Universität Berlin, Berlin, Germany.}
\affil[5]{Center for Ecological Research and Forestry Applications (CREAF), Campus de Bellaterra (UAB), Barcelona, Spain.}
\affil[6]{Department of Biology and Ecology of Fishes, Leibniz-Institute of Freshwater Ecology and Inland Fisheries, Berlin, Germany.}
\affil[7]{Faculty of Life Sciences, Albrecht Daniel Thaer-Institute of Agricultural and Horticultural Sciences, Humboldt Universit\"at zu Berlin, Berlin, Germany.}

\begin{document}
\onecolumn

\maketitle

\begin{abstract}
The collective dynamics and structure of animal groups has attracted the attention of scientists from different disciplines. A variety of agent-based models has been proposed to account for the emergence of coordinated collective behavior from simple interaction rules. A common, simplifying assumption of such collective movement models, is the consideration of individual agents moving with a constant speed. In this work we critically re-asses this assumption underlying a vast majority of collective movement models. First, we discuss experimental data showcasing the omnipresent speed variability observed in different species of live fish and artificial agents (RoboFish). Based on theoretical considerations accounting for inertia and rotational friction, we derive a functional dependence of the turning response of individuals on their instantaneous speed, which is confirmed by experimental data. We investigate how the interplay of variable speed and speed-dependent turning affects self-organized collective behavior by implementing an agent-based model which accounts for both effects. We show, that besides average speed, the individual speed variability may have a dramatic impact on the emergent collective dynamics, as two groups differing only in their speed variability, and being otherwise identical in all other behavioral parameters, can be in two fundamentally different stationary states (polarized versus disordered).
We find that the local coupling between group polarization and individual speed is strongest at the order-disorder transition, and that, in contrast to fixed speed models, the group's spatial extent does not have a maximum at the transition. Furthermore, we demonstrate a decrease in polarization with group size for groups of individuals with variable speed, and a sudden decrease in mean individual speed at a critical group size (N=4 for Voronoi interactions) linked to a topological transition from an all-to-all to a distributed spatial interaction network.
Overall, our work highlights the importance to account for fundamental kinematic constraints in general, and variable speed in particular, when modeling self-organized collective dynamics. 
\end{abstract}

\section{Introduction}
The emergent, highly coordinated, collective movements as observed in schools of fish, flocks of birds, or insect swarms, are fascinating examples of biological self-organization. Although our understanding of these collective systems has been significantly advanced over the past years through diverse research efforts in biology \cite{Berdahl2013, Katz2011a, Jolles2017, Ward2011, Krause1994a}, mathematics \cite{Torney2015, carrillo2010particle, ihle2011kinetic}, computer science \cite{Hidalgo2014, Olson2012}, engineering \cite{Li2020, Landgraf2016a}, and statistical physics \cite{vicsek2012collective, Cavagna2010, Feinermann, chate2008modeling, peruani2011traffic}, many fundamental questions remain open, for example regarding the underlying interaction networks ("Who interacts with whom?") \cite{Ballerini2008a,Strandburg-Peshkin2018,ling2019behavioural}.

In addition to the analysis of empirical observations \cite{Cavagna2010, Strandburg-Peshkin2018}, mathematical models are an important tool for studying self-organization and collective behavior, and have been instrumental in uncovering general principles of how robust, large-scale coordination can emerge from simple, local interactions between self-propelled agents \cite{vicsek1995novel,Olson2012, KlamserRomanczuk2021}.

When formulating models, in general and for animal collectives, one has to balance between simplicity/generality and detailed resemblance to experimental systems.
From a statistical physics point of view, it is viable to assume some sort of universality of the collective dynamics even in far-from-equilibrium situations.
Thus, as long as the model accounts for crucial aspects of the microscopic dynamics, other microscopic details become irrelevant for the macroscopic behavior for sufficiently large systems over a long temporal scale.
However,  1) there is no general way to tell when the system is sufficiently large, and 2) animal groups consist of tens to hundreds, rarely thousands or more, individuals. Therefore, animal collectives should be rather viewed as mesoscopic systems, where the actual details of individual movement behavior may play an important role \cite{Jolles2020tree}, and caution is advised when simplifying modeling assumptions.

A particularly prominent simplification often encountered in flocking models is the assumption of constant speed of individual agents, used in a vast majority of flocking models \cite{Couzin2002, vicsek2012collective, Gautrais2012, Jhawar2020}; for exceptions see e.g. \cite{Grossmann2012, Mishra2012,Harpaz2017, Calovi2018, SbragagliaKlamser2020}. However, in general animals are able to flexibly modify their movement speed. Thus, speed adaptation due to environmental factors or social interactions \cite{Katz2011a} -- ignored in constant speed models -- may play a decisive role in shaping the ability of groups to coordinate their movement and in the resulting structure of moving animal groups \cite{Kent2019}.
In fact, experiments demonstrated that speed influences the collective behavior strongly, via a coupling to polarization/alignment \cite{Mishra2012, Grossmann2012, Gautrais2012, Jolles2017, Jolles2020, Kent2019} which could also be shown on the local scale \cite{Mishra2012}, i.e. regions in the shoal with faster fish are more polarized.
While in former simulation studies the speed influenced the turning rate or modified the assumed social forces, mostly fixed speed models were used rendering the speed to a mere parameter and not a changing variable \cite{Couzin2002, Gautrais2012, Calovi2014, Jolles2017}.
It has been shown that a variable speed of individuals may lead to qualitatively new, emergent phenomena on the group level as for example bi-stable behavior with respect to polarization, i.e. a co-existence of stable ordered states and disordered collective states at high densities \cite{Grossmann2012, Mishra2012}.
These findings demonstrate the important role of feedbacks between speed, turning, and social interactions for the emergence and stability of collective states. 
Here, in the context of the collective movement of self-propelled agents that are meant to represent animal groups in the real world, we will discuss how inertia and friction link those properties.

In general, individuals in a group can differ in persistent behavioral traits (animal personality) which may induce between-individual variance in speed \cite{Reale2007, Jolles2017, Bierbach2017, Herbert-Read2013}).
The inter-individual variability in preferred movement speed has been found to influence cohesion and polarization of groups \cite{Jolles2017, Jolles2020}, and it has been shown to decrease in larger groups \cite{Herbert-Read2013}. However, already in behaviorally homogeneous groups, with individuals having similar preferred speeds, the instantaneous speed of single individuals will dynamically vary over time due to its direct response to social and/or environmental cues, as well as due to internal decision processes ("internal" fluctuations). Both types of speed variability will be important for the collective movement dynamics \cite{Grossmann2012}. We focus on the investigation of the role of within-individual speed variability on emergent, self-organized collective movement using an agent-based model. In particular, we demonstrate how accounting for the ability of individuals to dynamically modulate their speed has profound effects on the group behavior, which are highly relevant for experimental observations of collective animal movement.

In the following, we will first provide an experimental motivation for our modeling ansatz by analyzing speed variability in schooling fish and providing evidence for coupling between turning behavior and instantaneous speed, which can be theoretically understood by considering self-propelled movement with inertia. Inspired by these results we will then investigate an agent-based model of collective movement with variable speed and demonstrate how the ability of individuals to flexibly adapt their speed in response to social interactions and fluctuations has major consequence for the emergent collective dynamics.  

\section{Methods}
\subsection{Experimental Data} 
In order to show within-individual variability in  movement speeds, we analyzed previously published data sets of individual movement for two different fish species \textit{Poecilia reticulata} (Trinidadian guppy, \cite{Jolles2020}) , \textit{Poecilia formosa} (clonal Amazon molly, \cite{Bierbach2017})) as well as a biomimetic robot ('RoboFish', \cite{Jolles2017}).
For the Amazon molly, we further included a data set in which groups of 4 fish were observed \cite{Doran2019}.
All data sets consist of positional tracking data from laboratory observations with a sampling frame rate of 30 fps, circular or rectangular arenas smaller than 1 square meter in size and only female fish, as summarized in Tab.~\ref{tab:experimental_info} and in more detail explained in SI Sec.~\ref{sec:SI_expSetups}.
\begin{table}[h]
\centering
\begin{tabular}{ l c c c c }
             & RoboFish           &  Guppy         & Molly (single ind.) & Mollies (groups of 4) \\
             \cmidrule[1pt](lr{.75em}){2-2}
             \cmidrule[1pt](lr{.75em}){3-3}
             \cmidrule[1pt](lr{.75em}){4-4}
             \cmidrule[1pt](lr{.75em}){5-5}
Species             & -- & \textit{Poecilia reticulata} & \multicolumn{2}{c}{\textit{Poecilia formosa}} \\
\# of tracks      & 39                & 40                 & 35                      & 32                \\ 
Observation time    & 10 min            & 10 min            & 6 min                 & 5 min             \\
Arena dimensions   & 88$\times$88 cm   & 88$\times$88 cm   & 48.5cm diameter       & 60$\times$30 cm   \\
Water depth         & 7.5 cm            & 7.5 cm            & 3 cm                  & 5 cm              \\
Frame acquisition   & 30 FPS            & 30 FPS            & 30 FPS                & 30 FPS            \\
Sex                 &   --          &  female               & female                & female               \\
Tracking method                 & BioTracker \cite{Monck2018}        &  BioTracker        & Ethovision (10.1)     & Ethovision (XT12) \\
Reference           & \multicolumn{2}{c}{Jolles et al. \cite{Jolles2020}}& Bierbach et al. \cite{Bierbach2017}  & Doran et al. \cite{Doran2019}
\end{tabular}
\caption{\textbf{List of previously published tracking data used in our analysis.}
    The table lists major characteristics of the datasets we used to show within-individual speed variability. 
    The \# of tracks indicates the number of individual tracks used for the analysis. Du to the initial study designs and questions, tracks may represent repeated measures of the same (Guppy: 20 individuals, RoboFish 1 replica) or different individuals (Molly single ind, Molly groups: 8 with 4 ind. per group). Please find exact study designs in the respective references.
    }
\label{tab:experimental_info}
\end{table}
\subsection{Processing of trajectories}
The tracking data obtained for the different species and the robotic fish encodes the position $\mathbf{x}_i(t) = \big[ x_i(t), y_i(t) \big]^T$ of the individual $i$ for each frame $t$.
We approximate the velocity of each individual from subsequent positions by computing:
\begin{align}
    \mathbf{v}_{i,x}(t+\Delta t) = \frac{\mathbf{x}_i(t+ \Delta t) - \mathbf{x}_i(t)}{\Delta t}, \label{eq:speedX}\ .
\end{align}

We can approximate the direction of motion of individual $i$ by $\varphi_i(t) = \arctantwo{\big( v_{i,y}(t),\ v_{i,x}(t) \big)}$.
In a similar way as in Eq.~(\ref{eq:speedX}) we compute the angular speed $\dot{\varphi}(t)$ of each individual.

\subsection{Fundamental relations between speed and turning} \label{sec:SpeedAndTurning}

The fundamental equation of motion for a self-propelled agent $i$ reads: 
\begin{align}
    \frac{\text{d} \mathbf{v}_i(t)}{\text{d} t} = \frac{1}{m}\mathbf{F}_i(t)
\end{align}
with $\mathbf{v}_i$ as the velocity vector of the agent, $m$ it's mass and $\mathbf{F}_i$ being total  force acting on it. 
Please note that, in the following, we omit the explicit time dependence for simplicity.
The velocity vector can be expressed via the speed $v_i$ and the heading angle $\varphi_i$ to $\mathbf{v}_i = v_i [\cos \varphi_i, \sin \varphi_i]^T = v_i\ \hat{\mathbf{e}}_{v, i}$ . 
We can reformulate (in detail shown in SI Sec.~\ref{sec:SI_deriveTurning}) the velocity dynamics in terms of speed and heading angle dynamics \cite{romanczuk2012active} to 
\begin{align}
    \frac{\text{d} v_i}{\text{d} t} &= 
    \frac{\mathbf{F}_i}{m} \cdot \hat{\mathbf{e}}_{v, i} \label{eq:dvdt}\\
    \frac{\text{d} \varphi_i}{\text{d} t} &= 
    \frac{\mathbf{F}_i}{v_i\ m} \cdot \hat{\mathbf{e}}_{\varphi, i}\ \text{ with }\  \hat{\mathbf{e}}_{v, i} = \begin{bmatrix}\cos \varphi_i \\
                                         \sin \varphi_i\end{bmatrix}\ .
    \label{eq:dphidt}
\end{align}
Therefore, without any further assumptions, we see that the turning is inversely proportional to the current speed, i.e. $\text{d} \varphi_i / \text{d} t \propto 1/v_i$. 
However, the inverse proportionality results in instantaneous turning for $v_i=0$, which is unrealistic and is caused by assuming a point-like object.
To provide a simple correction for this unreasonable assumption, we follow \cite{Romanczuk2011_diss} and introduce a rotational friction force acting on the velocity
\begin{align}
    \frac{\text{d} \mathbf{v}_i}{\text{d} t} = \frac{1}{m}\left( \mathbf{F}_i  - \alpha\  \frac{\text{d} \varphi_i}{\text{d} t}\ \hat{\mathbf{e}}_{\varphi, i} \right) \label{eq:dvdt_2}
\end{align}
with $\alpha$ as rotational friction coefficient.
If we repeat the steps from above analogously, the speed dynamics remain unchanged (Eq.~\ref{eq:dvdt}) but the change in heading angle reads now:
\begin{align}
    \frac{\text{d} \varphi_i}{\text{d} t} = 
    \frac{\mathbf{F}_i}{(v_i + \alpha)\ m} \cdot \hat{\mathbf{e}}_{\varphi, i}\ . \label{eq:dphidt_2}
\end{align}

In the context of self-propelled agents, the above relation implies that the turning rate of an individual, in response to a force $\mathbf{F}_i$ acting on the agent or generated by the agent itself, depends on its speed $v_i$. For a constant force $|\textbf{F}_i|=const.$, faster agents will turn slower. Alternatively, in order to turn at the same rate, individuals moving at different speeds have to adjust the strength of their turning force linearly with their current speed.
We  emphasize that this fundamental relation, ignored in most models of collective behavior explicitly modeling turning rates, holds both for fixed speeds $v_i:=v_{0,i}=const.$, as well as for variable speeds $v_i:=v_i(t)$.  
\subsection{Fitting experimental data}
We used the least square fitting method to obtain the best fitting functional dependency between $v=|\mathbf{v}|$ and $\dot{\varphi}$. 
%
We considered the two equations (\ref{eq:dphidt}) and (\ref{eq:dphidt_2}), introduced in the preceding Sec.~\ref{sec:SpeedAndTurning}, and thus minimizing the squared difference of the data to $\text{d}\varphi / \text{d} t = F_{\varphi}/v$ and $\text{d}\varphi / \text{d} t = F_{\varphi}/(v+\alpha)$ respectively.
Here we treat the force in angular direction $F_{\varphi} = \mathbf{F} \cdot  \hat{\mathbf{e}}_{\varphi}$, and the rotational friction coefficient $\alpha$ as parameters.
Since Eq.~\ref{eq:dphidt_2} has one parameter more ($\alpha$), we compared both fits using the Akaike Information Criterion (AIC) \cite{Akaike1974} and the Bayesian Information Criterion (BIC) \cite{Schwarz1978}.
Those criteria penalize a larger number of parameters and therefore prevent overfitting.

\subsection{The Model} \label{sec:theModel}

\begin{figure}
    \centering
    \includegraphics[width=\textwidth]{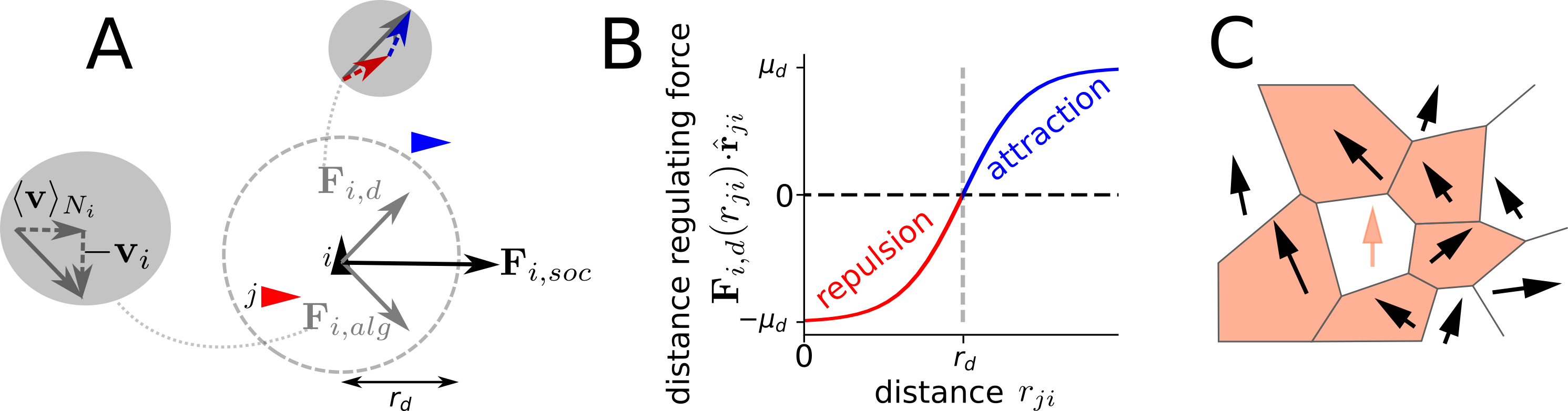}
    \caption{
        \textbf{Implementation of social interactions.}
        A focal agent (black triangle) responds to neighbors (red and blue triangle) via a velocity alignment force $\mathbf{F}_{i, alg}$, which aims at minimizing the velocity difference to the mean neighbor velocity $\langle \mathbf{v} \rangle_{N_i}$, and a distance regulating force $\mathbf{F}_{i, d}$, their sum corresponds to the social force $\mathbf{F}_{i, soc}$(\textbf{A}).
        For simplicity the distance regulating force has a sigmoidal distance dependence: it is repulsive for distances closer, and attractive for distances larger than a preferred distance $r_d$ (\textbf{A, B}).
        The neighborhood of a focal agent (red arrow) is defined by its Voronoi neighbors (black arrows in red cells, \textbf{C}).
    }
    \label{fig:model}
\end{figure}
As explained in section \ref{sec:SpeedAndTurning}, the model we use mimics the movement behavior of real fish by obeying fundamental physics relations (inertia and friction).
This is mathematically expressed in Eqs.~\ref{eq:dvdt_2}, \ref{eq:dphidt_2}.
Additionally, the interaction between fish is motivated by the continuous version of a well established three zone model (already introduced with fixed speed in \cite{KlamserRomanczuk2021}).
Thus, the force acting on an individual $i$ has a self-propulsion term (including noise) and a social term.
We can express this as: $\mathbf{F}_i(t) =  \mathbf{F}_{i, \text{sp}}(t) + \mathbf{F}_{i, \text{social}}(t)$.
The self-propulsion force takes into account two main factors: 1) the tendency of an individual to keep a preferred speed $v_0$ and 2) the fluctuations on the linear speed $v$ and the angular speed $\dot{\varphi}$
\begin{align}
    \mathbf{F}_{i, \text{sp}}(t) = \bigg( \beta \big( v_0 - v_i(t) \big) + \sqrt{2 D_v} \, \xi_{v}(t) \bigg) \mathbf{\hat{e}}_{v,i} + \bigg( \sqrt{2 D_{\varphi}} \, \xi_{\varphi}(t) \bigg) \mathbf{\hat{e}}_{\varphi,i},
    \label{eq:selfpropForce}
\end{align}
where $\beta$ is the speed relaxation coefficient, leading to the relaxation of the speed towards the preferred speed $v_0$ in the absence of other perturbations with the time constant $\tau_v=\beta^{-1}$.  $D_v$ and $D_{\varphi}$ are diffusion coefficients setting the noise intensity in $v$ and $\varphi$, respectively, whereas $\xi_v$ and $\xi_{\varphi}$ are independent, Gaussian white noise processes. 
The social interactions are explained in detail in the following.
\subsubsection{Social interactions}
We consider a social force that combines two fundamental types of interactions among individuals: 1) an alignment force $\mathbf{F}_{i,alg}$ and 2) a distance-regulating force $\mathbf{F}_{i,d}$ (Fig.~\ref{fig:model}A, B).
Thus, we can express the total social force as $\mathbf{F}_{i,\text{social}}(t) = \mathbf{F}_{i,alg}(t) + \mathbf{F}_{i,d}(t)$.
We use Voronoi tesselation to define the neighborhood of a focal individual $i$, which is labeled as $\mathbb{N}_i$ (Fig.~\ref{fig:model}C). A Voronoi interaction network can, on the one hand, be efficiently computed, while on the other hand it also shows a good approximation with visual interaction networks \cite{strandburg2013visual}. 
The mathematical expression of the alignment force is:
\begin{align}
    \mathbf{F}_{i,alg}(t) = \frac{1}{|\mathbb{N}_i|} \sum_{j \in \mathbb{N}_i} \mu_{alg} \, \mathbf{v}_{ji}(t),
    \label{eq:AlignmentForce}
\end{align}
where $\mu_{alg}$ is the alignment strength and $\mathbf{v}_{ji}(t) = \mathbf{v}_j(t) - \mathbf{v}_i(t)$.
The distance-regulating social force assumes a preferred distance $r_d$ that individuals try to maintain between  each other. It is defined as:
\begin{align}
    \mathbf{F}_{i, d}(t) = \frac{1}{|\mathbb{N}_i|}\sum_{j \in \mathbb{N}_i} \mu_d \ \tanh{\bigg(m_d \big(r_{ji}(t) - r_d \big)\bigg)} \ \mathbf{\hat{r}}_{ji}(t),
    \label{eq:RepAttrackForce}
\end{align}
where $\mathbf{\hat{r}}_{ji} = (\mathbf{r}_j - \mathbf{r}_i)/|\mathbf{r}_j - \mathbf{r}_i|$ is a unitary vector from agent $i$ to agent $j$, $r_{ji} = |\mathbf{r}_j - \mathbf{r}_i|$, $\mu_d$ is the strength of the force and $m_d$ is the slope of the change from repulsion ($r_{ji} < r_d$) and attraction ($r_{ji} > r_d$) (Fig.~\ref{fig:model}B).
In principle, it is possible to extract a specific functional form of the repulsion and attraction interaction from experimental data \cite{Katz2011a, Herbert-Read2011, Calovi2018}. However, these functions will likely depend on the species and the ecological context, whereas the qualitative role of variable speed discussed below does not depend on the specific choice of the functional form of the inter-individual attraction-repulsion interactions.   
Therefore, for the sake of simplicity and generality, we have chosen a rather simple (sigmoidal) distance dependence for the distance regulating force controlled by only three parameters ($\mu$, $m_d$, $r_d$), with the key property being a finite preferred distance $r_d$, which individuals try to keep to their neighbors. 

\subsubsection{The equations of motion}
By considering the self-propulsion and social forces described above, we can write the explicit equations of motion for individuals, which resemble the equations in \cite{Grossmann2012}:
\begin{align}
    \frac{\text{d}v_i(t)}{\text{d}t} &= \beta \big(v_0 - v_i(t)\big) + F_{i, v}(t) + \sqrt{2 D_{v}} \,\, \xi_v(t)\label{eq:EqMotVarySpeed_2}\\
    \frac{\text{d}\varphi_i(t)}{\text{d}t} &= \frac{1}{v_i(t) + \alpha}\left(F_{i, \varphi}(t) + \sqrt{2 D_{\varphi}} \,\, \xi_{\varphi}(t) \right),\label{eq:EqMotVarySpeed_3}
\end{align}
with $F_{i, v}(t) = \mathbf{F}_{i, \text{social}}(t) \cdot  \mathbf{\hat{e}}_{v,i}(t)$ being the projection of the social force on the heading direction $\mathbf{\hat{e}}_{v,i}$ and $F_{i, \varphi}(t) = \mathbf{F}_{i,\text{social}}(t) \cdot \mathbf{\hat{e}}_{\varphi, i}(t)$ being the projection of the social force on the turning direction $\mathbf{\hat{e}}_{\varphi, i}$.

\section{Results}
\subsection{Experimental Data of Individual Fish}
\begin{figure}[]
    \begin{center}
    \includegraphics[width=1\textwidth]{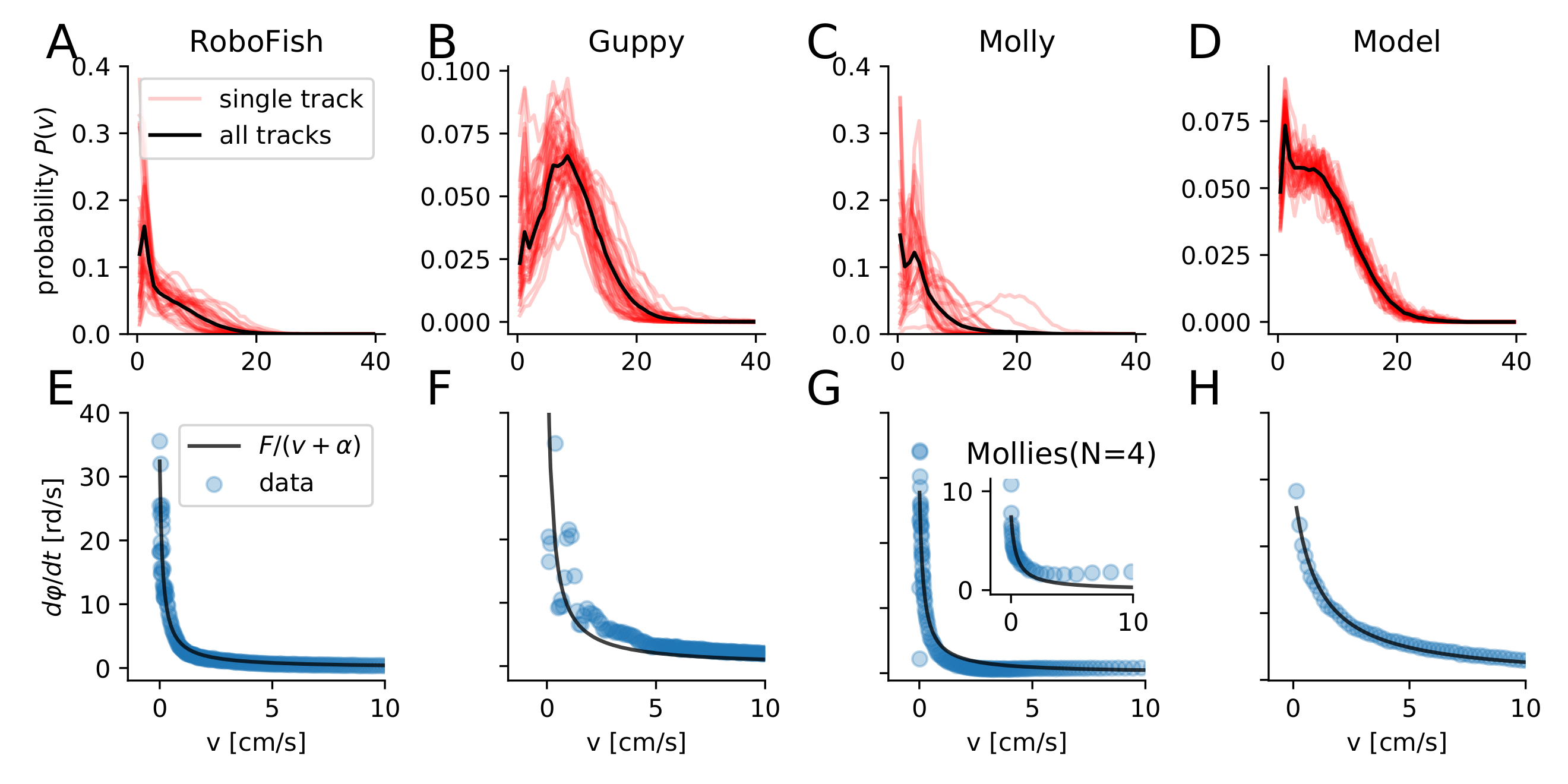}
    \end{center}
    \caption{
      \textbf{Speed and turning of individuals.}
      The speed probability distributions $P(v)$ for the experimental data (RoboFish, guppy, molly) and for model simulations of individuals (\textbf{A-D}). Red transparent lines represent single tracks and the solid black line is the distribution of all tracks summarized. 
      The absolute turning rate $\dot{\varphi}$ as a function of the speed $v$ for the different individuals(\textbf{E-H}).
      For mollies we also show the relation between turning rate speed of individuals swimming in groups of $N=4$ (inset \textbf{G}).
      The parameters of the model simulations (described in Section~\ref{sec:theModel}) are listed in Tab.~\ref{tab:paraSummary} and those of the model fits in Tab.~\ref{tab:modelFitParas}.
    }
    \label{fig:expEvidence}
\end{figure}
The speed variability and the dependence of turning on the instantaneous speed are fundamental characteristics of the movement of individual self-propelled agents (c.f. \cite{peruani2007self,Romanczuk2011_diss}). In Fig.~\ref{fig:expEvidence}A-C,E-G, we show the corresponding experimental results obtained from analyzing the trajectories of two different fish species (guppies \& mollies) and the robotic fish. We show that live as well as artificial agents exhibit the same qualitative behavior: 1) non-negligible speed variability and 2) decrease of turning rate with increasing speed. The latter, in particular, can be linked to fundamental kinematic constraints of inertial motion discussed above.  
We apply the same analysis on trajectories of our model simulation, which produces the same characteristics Fig.~\ref{fig:expEvidence}D,H.

We find for all cases a speed distribution that shows a strong variation in speed (Fig.~\ref{fig:expEvidence}A-D). The Coefficient of variation ($\text{COV}(v) = \sigma_v/\langle v \rangle$) of the speed is about 1, i.e. the speed variation is as large as the mean speed. 
The results of the single track analysis are: $\text{COV}(v) =  0.92 \pm 28$   ({RoboFish}), $0.57 \pm 0.1$ ({guppy}), $0.92 \pm 0.78$  ({single molly}), $0.68 \pm 0.03$  ({model}).

Thus, speed variation typically neglected by fixed speed models, is clearly evident in the experimental data and accounted for in our variable speed model, Fig.~\ref{fig:expEvidence}D.
Accounting for variable speed is important, as due to inertia the turning ability of any object is inversely proportional to its speed $d \varphi / d t = F/v$.
If we take turning friction into account (friction coefficient $\alpha$, see Methods), the turning rate becomes $d \varphi / d t = F/(v+\alpha)$. 
The inverse speed dependence of the turning rate is observable for all 4 cases (Fig.~\ref{fig:expEvidence}E-H).
A least square fit of the two turning rate models and their comparison via the Akaike- (aic) and Baysian Information Criterion (bic) suggests that the model that additionally takes turning friction into account explains all data sets best (Fig.~\ref{fig:expEvidence}E-H and Tab.~\ref{tab:modelCompare}).
The same holds for individuals swimming in groups of fish (mollies $N=4$, inset Fig.~\ref{fig:expEvidence}H).

Note that in our variable speed model the dependence of the turning rate on speed and turning friction is hard coded.
Thus, the individual turning of simulated agents resembles qualitatively (in terms of the functional dependence) the behavior of real fish.
This enables us to explore how social interactions in combination with variable speed and turning restriction affect collective behavior.

\begin{table}[]
  \centering
    \begin{tabular}{l cc cc cc cc cc}
                                            & \multicolumn{2}{c}{RoboFish} & \multicolumn{2}{c}{Guppy} & \multicolumn{2}{c}{Molly} & \multicolumn{2}{c}{Mollies(N=4)} & \multicolumn{2}{c}{Model} \\
                                    \cmidrule[1pt](lr{.7em}){2-3}      \cmidrule[1pt](lr{.75em}){4-5}   \cmidrule[1pt](lr{.75em}){6-7}\cmidrule[1pt](lr{.75em}){8-9}\cmidrule[1pt](lr{.75em}){10-11}
                                                   & aic & bic                 & aic & bic                    & aic & bic               & aic & bic                 & aic & bic\\
    $\frac{d\varphi}{dt}= F/v$                     & 71 & 72                 & 1170 & 1173                        & 595 & 598           & 73 & 75                   & 229 & 231 \\
    $\frac{d\varphi}{dt}= F/(v+\alpha)$            & -24 & -22               & 523 & 530                        & 316 & 322             & 11 & 15                  & -107 & -102 \\
    \end{tabular}
  \caption{\textbf{Statistical model comparison.} Akaike (aic) and Baysian (bic) information criterion for a model without ($\frac{d\varphi}{dt}= F/v$) and with ($\frac{d\varphi}{dt}= F/(v+\alpha)$) turning-friction $\alpha$ for each model-species. The values of the parameters $\alpha$ and $F$ are listed in Tab.~\ref{tab:modelFitParas}.
  }
  \label{tab:modelCompare}
\end{table}

\subsection{Collective level consequence of speed variability}

We have shown so far that large speed variability is a common feature of live and robotic fish's moving pattern and that the individual turning rate strongly depends on the current speed.
Importantly, our agent-based model, accounting for inertia and (rotational) friction, reproduces the corresponding characteristics. This allows us to use now our mathematical model to systematically explore the impact of these movement characteristics at the collective level in groups with $N=400$, corresponding to large schools of fish in the wild.
\subsubsection{Order induced by speed and speed-variation}
\begin{figure}[]
    \includegraphics[width=1\textwidth]{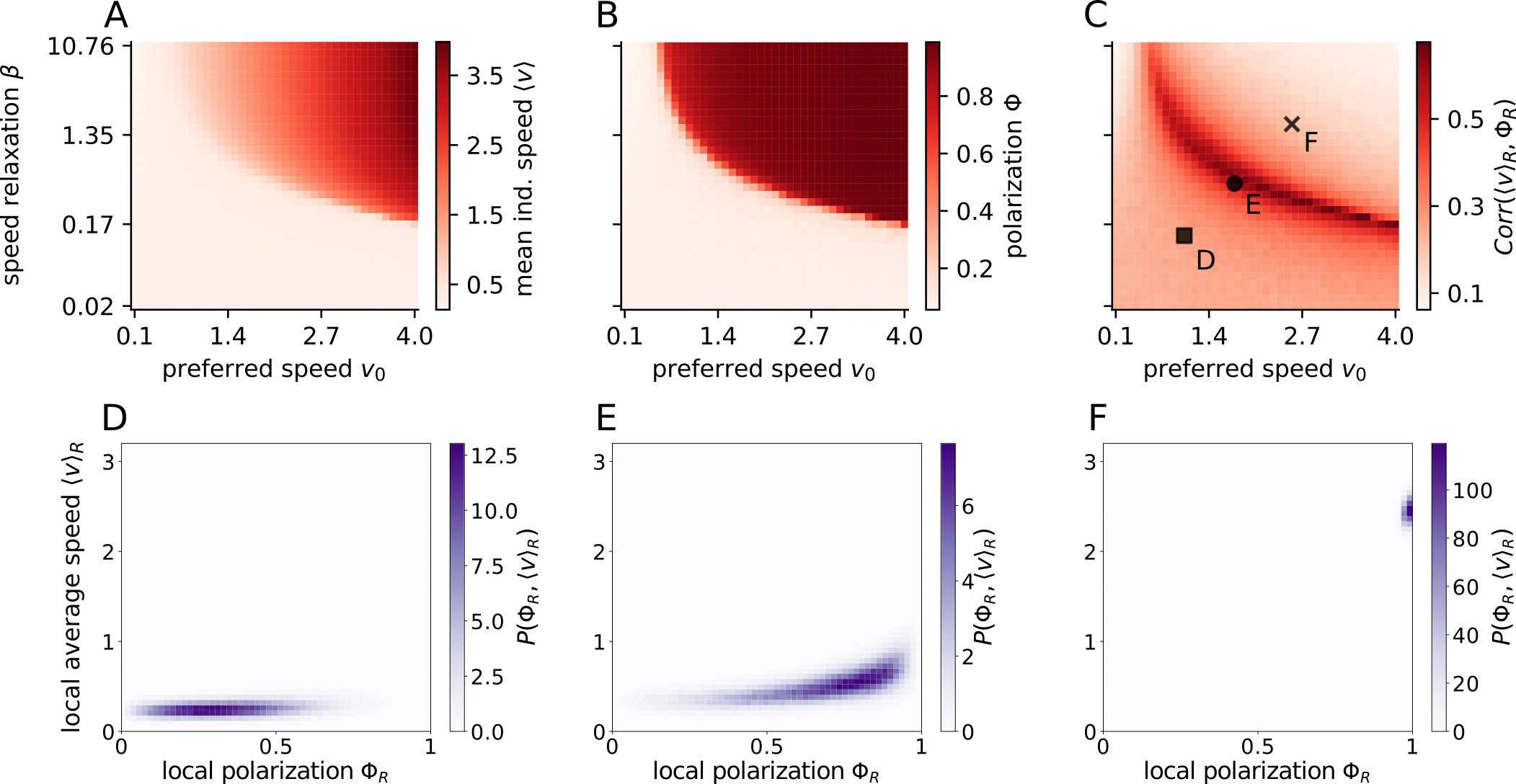}
    \caption{
    \textbf{Influence of preferred speed and speed variability on polarization and speed-polarization coupling.}
    The preferred speed $v_0$ and the speed relaxation strength $\beta$ both affect the individual speed (\textbf{A}) and modulation of either of these parameters may induce orientational order marked by a high polarization $\Phi$ (\textbf{B}).
    The transition to order can be understood by a local coupling between the local average speed $\langle v \rangle_R$ and the local polarization $\Phi_R$, which we quantified via their correlation $Corr(\langle v \rangle_R,\ \Phi_R)$  (\textbf{C}).
    The local averages consider all individuals within a circle of radius $R=3$ around a focal one.
    Note that $\langle v \rangle_R$ is not the local group velocity (where a positive correlation with order is trivial) but the local average of the individual speed magnitudes.
    The specific dependence between local speed and order for three specific parameter choices (marked by square, circle and cross in \textbf{C}) is shown for the disordered state (\textbf{D}), the phase transition region (\textbf{E}) and the ordered state (\textbf{F}).
    The parameters in the simulations are listed in Tab.~\ref{tab:paraSummary}.
    }
    \label{fig:PhaseTransition}
\end{figure}
Animals can vary in their preferred speed $v_0$ and also in their within-individual speed variability, which is parametrized by the speed relaxation coefficient $\beta$.
For socially interacting agents, the mean individual speed $\langle v\rangle$ is close the preferred speed $v_0$ only in the ordered state (Fig.~\ref{fig:PhaseTransition}A, B).
Interestingly, a group can be ordered and another that is identical in all parameters except in the preferred speed and/or the speed variability can be in a disordered state.
As shown for real, robotic and simulated fish (Fig.~\ref{fig:expEvidence}E-H) the turning is slower the higher the speed.
This causes rotational random forces to be damped for groups with larger speeds, facilitating order due to inertial restrictions on turning (Fig.~\ref{fig:PhaseTransition}B).

In contrast, large speed-variability (low $\beta$) may lead to disorder, while a narrow individual speed distribution (large $\beta$) induces order.
If the speed of an agent can vary (low relaxation coefficient $\beta$) the velocity alignment can effectively reduce the average speed of individuals:
A focal agent $i$ aligns with the mean velocity of its neighbors $\langle \mathbf{v} \rangle_{\mathbb{N}_i}$. However, for finite levels of directional fluctuations $|\langle \mathbf{v} \rangle_{\mathbb{N}_i} | \lessapprox v_i$, i.e. it will decelerate due to an effective social friction associated with the alignment interaction \cite{Grossmann2012}.
The reduced speed allows a faster turning and consequently enhances the angular noise and therefore disorder (Fig.~\ref{fig:PhaseTransition}B).
Thus, in any collective system in which individuals align velocity vectors and not only orientations, i.e. try to match not only the direction of motion but also their speed to the average perceived movement of the local neighborhood, different collective states can emerge due to individuals in the corresponding groups only differing in their speed variability with all other behavioral parameters being identical.

The dynamic speed variability (low $\beta$) has another highly robust emergent consequence.
It allows agents of the same collective to differ in their instantaneous speed and since higher speeds induce order, we observe correlations on the local level between mean individual speed $\langle v \rangle_R$ and local polarization $\Phi_R$ with $R$ as the radius of the circle from which the average is computed (Fig.~\ref{fig:PhaseTransition}C-F). Please note that as we consider individual speed, the above correlation is different from the trivial correlation between local polarization and local group speed.
The correlations between individual speed and local polarization is always positive and largest at the transition between disorder and order.
The latter is a signature of second order phase transitions, where the susceptibility, i.e. the response to weak signals/fluctuations, is maximal.
It means that information encoded in speed is best translated to a directional response at the transition region, and vice versa (likely to be beneficial in collective computation tasks).

The local coupling is an emergent consequence of the fundamental dependence of turning on speed.
Thus, it is highly robust and the qualitatively same non-linear functional form was observed in experiments (compare Fig.~\ref{fig:PhaseTransition}D-F with Fig.~1 in \cite{Mishra2012}).
Most importantly, it weakens with low speed variability (Fig.~\ref{fig:PhaseTransition}C) and does not exist for fixed speed models.

\subsubsection{Group structure and speed}

\begin{figure}
    \includegraphics[width=1\textwidth]{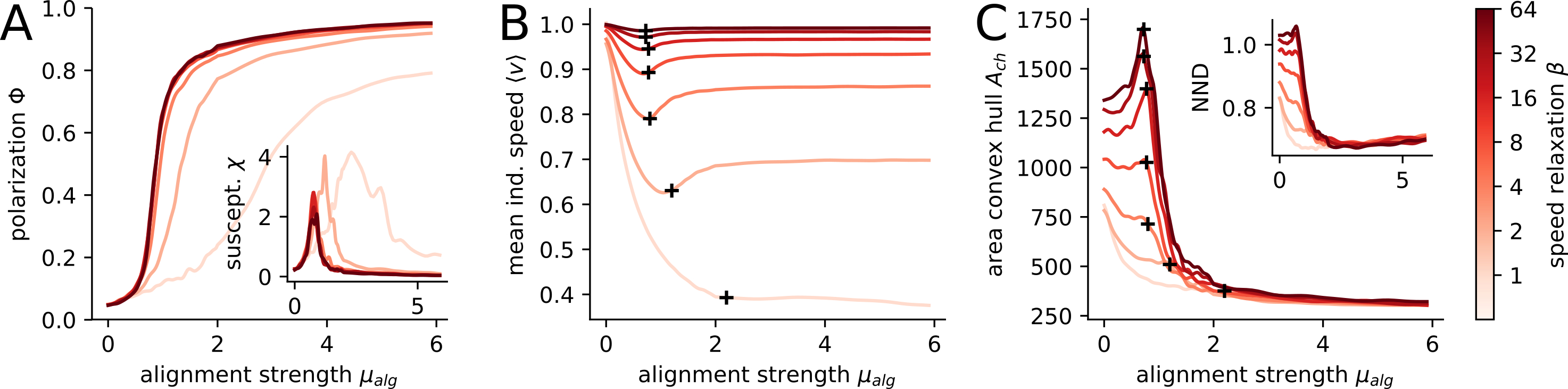}
    \caption{
    \textbf{Variable speed affects collective behavior in different states.}
    Polarization $\Phi$(\textbf{A}), the susceptibility $\chi$ (inset \textbf{A}), the individual speed $\langle s\rangle$ (\textbf{B}), the area of the convex hull (\textbf{C}) and the nearest neighbor distance $NND$ (inset \textbf{C}) are shown in dependence on the velocity alignment strength $\mu_{alg}$.
    Black crosses (\textbf{B, C}) mark the peak of the susceptibility, i.e. the location of the phase transition.
    The lines are color coded according to the speed relaxation strength as indicated at the colorbar, i.e. $\beta\in\{1,\ 2,\ 4,\ 8,\ 16,\ 32,\ 64\}$.
    }
    \label{fig:GroupStructure}
\end{figure}

We have demonstrated above that the preferred speed and its variability can induce an order-disorder transition.
Now, we keep the preferred speed fixed at $v_0=1$ and change the alignment strength $\mu_{alg}$.
By repeating this for different speed relaxation strength $\beta$ we investigate how the collective behaves in the ordered and disordered state (controlled by $\mu_{alg}$) depending on how variable the speed is.

The higher the speed variability of individuals, the larger alignment strengths are necessary for the collective to reach the ordered state (Fig.~\ref{fig:GroupStructure}A).
The shift of the phase transition is more clearly depicted by shifting peaks of the susceptibility $\chi = N (\langle \Phi ^2 \rangle - \langle \Phi \rangle^2)$ (fluctuations of the polarization, Fig.~\ref{fig:GroupStructure}A inset).

The collective phase transition impacts the individual dynamics as well.
The mean individual speed $\langle s \rangle$ shows a distinct minimum at the transition which vanishes for low speed relaxation strength $\beta=1$ (Fig.~\ref{fig:GroupStructure}B).
The minimum in speed is related to the velocity alignment where a focal agent adjusts its velocity $\mathbf{v}_i$ to the average velocity vector of its neighbors $\langle \mathbf{v} \rangle_{\mathbb{N}_i}$.
In the disordered state $\langle \mathbf{v} \rangle_{\mathbb{N}_i} \approx \mathbf{0}$, i.e. the alignment interaction induces an effective social friction $-\mu_{alg} v_i$ and thus slows the focal agent down \cite{Grossmann2012}.
It changes at the disorder-order transition where the neighborhood of each agent becomes increasingly polarized with increasing alignment strength.
However, since there is always noise on the heading direction $|\langle \mathbf{v} \rangle_{N_i}| < v_0$, even in the strongly ordered state the individual speed is below the preferred speed $v_0$.

A very general qualitative change from fixed to variable speed can be observed with respect to group structure close to the phase transition.
The area of the convex hull of the collective is maximal at the transition for fixed speeds.
This maximum becomes less pronounced and finally vanishes with increasing speed variability (lower speed relaxation strength $\beta$, Fig.~\ref{fig:GroupStructure}C).
The same holds for the nearest neighbor distance (Fig.~\ref{fig:GroupStructure}C inset).
At the transition the directional correlation of the agents is maximal (i.e. susceptibility peaks, Fig.~\ref{fig:GroupStructure}A inset) and the directional fluctuations cause subgroups of the collective to head in different directions, leading to an expansion of the collective \cite{KlamserRomanczuk2021}.
This expansion weakens with increasing speed variability because the distance regulating force can now lower the speed from a subgroup if it moves away from the shoal, effectively inhibiting expansion.

\subsubsection{Group size dependent effects}
%
\begin{figure}[h]
    \begin{center}
    \includegraphics[width=1\textwidth]{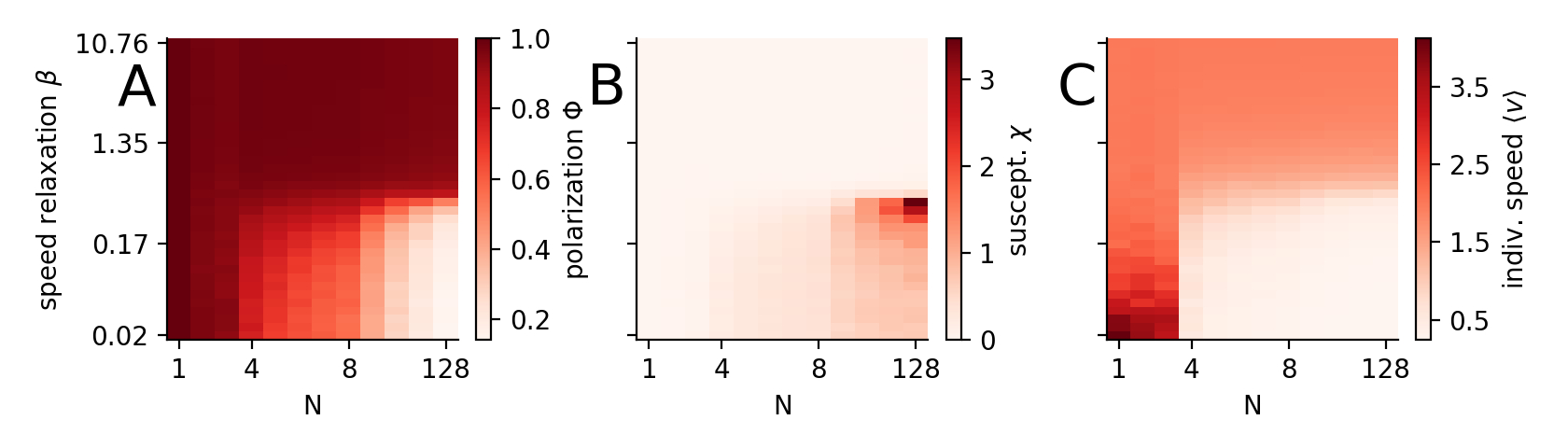}
    \end{center}
    \caption{
    \textbf{Group size effects depend on speed variability.}
    The polarization $\Phi$ (\textbf{A}), the susceptibility $\chi$ (\textbf{B}) and the mean individual speed $\langle v \rangle$ (\textbf{C}) are shown for different speed relaxation strength $\beta$ and group sizes $N$.
    }
    \label{fig:GroupSize}
\end{figure}
Group size is among the most biologically most important and experimentally most easily controllable parameters in the context of flocking and schooling. 
Thus, we investigate in this last part how group and individual measures change with group size $N$.

For high speed variability, i.e. low speed relaxation strength $\beta$, polarization decreases with increasing groups size and we expect $\Phi\to 0$ for even larger $N$ (Fig.~\ref{fig:GroupSize}A). For low speed variability (large $\beta$), the polarization remains high $\Phi \lesssim 1$ independent on $N$.
Note that only in a narrow range close to the transition (marked by a large susceptibility, Fig.~\ref{fig:GroupSize}B), the polarization saturates to intermediate values for large groups.

With group size being a key parameter, the question regarding existence of a threshold size, where the system's behavior changes qualitatively, is of particularly relevance.
Our results show that for agents with high speed variability (low $\beta$), the mean individual speed $\langle v \rangle$ undergoes a sudden change at $N=3$ (Fig.~\ref{fig:GroupSize}C).
Until $N=3$ the individual speed $\langle v \rangle$ is larger than $v_0$ and saturates towards $v_0$ with decreasing speed variability (increasing $\beta$).
The reason for $\langle v \rangle > v_0$ is that the speed distribution of individuals is asymmetric, with a long-tail at large speeds but cutoff at low speeds at $v=0$ (Fig.~\ref{fig:expEvidence}, A-D), i.e. a maximum of the distribution is at $v=v_0$ but the mean is larger.
For larger groups with $N\geq 4$, the speed is lower than the preferred speed but saturates also to $v_0$ in the fixed speed limit ($\beta\to\infty$).

\begin{figure}[]
    \begin{center}
    \includegraphics[width=1\textwidth]{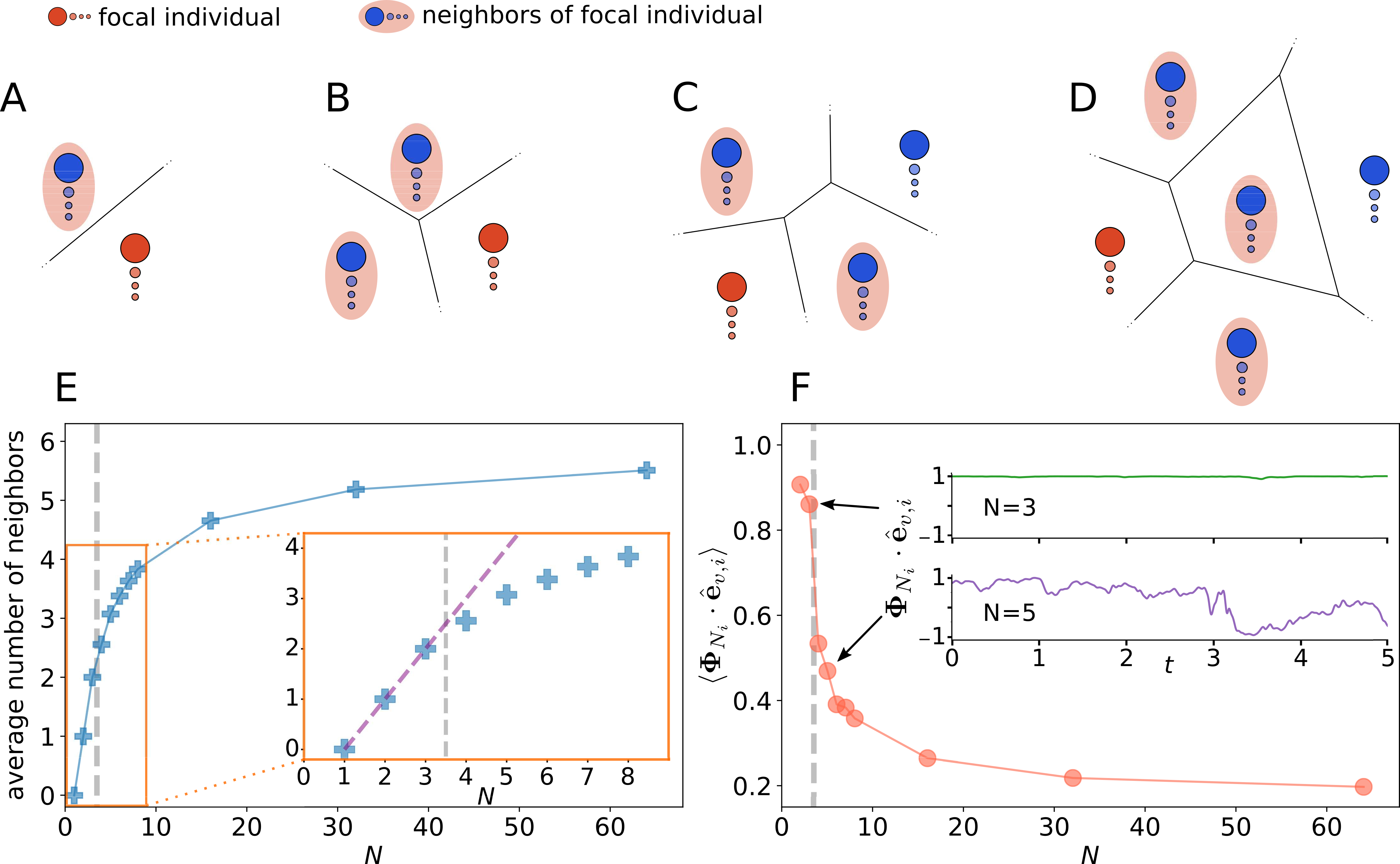}
    \end{center}
    \caption{
    \textbf{Qualitative topological change with group size.}
    \textbf{A-D}: illustrations of typical spatial constellations for different group sizes. The focal (red) agent and it's neighbors (blue in red ellipse) and for groups $N\geq4$ also agents that are not connected to the focal agent (blue) are shown. 
    \textbf{E}: average Voronoi neighbor number for different group sizes.
    The dashed purple line (inset \textbf{E}) marks the numbers of neighbors of a fully connected group.  
    \textbf{F}: average vector product of neighborhood polarization vector $\mathbf{\Phi}_{\mathbb{N}_i}$ and the heading direction $\hat{\mathbf{e}}_{v, i}$ of the focal agent $i$.
    The time series of the vector product (inset \textbf{F}) reveals a distinct difference between groups of $N=3$ and $N\geq4$.
    }
    \label{fig:GroupSizeExplained}
\end{figure}

This abrupt change can be understood through the interplay of individual dynamics and fundamental property of the interaction network: (i) A focal agent decelerates stronger the more its heading deviates from the average polarization of its neighborhood, i.e. $\text{d} v_i/ \text{d} t \propto \mathbf{\Phi}_{\mathbb{N}_i} \cdot \mathbf{\hat{e}}_{v,i} - 1$ (derived in SI Sec.~\ref{sec:SI_neighPolar}). (ii) For Voronoi-type interaction, for group size $N\leq 3$, we have an all-to-all interaction network, which is not the case for $N>3$.
The second point is illustrated in Fig.~\ref{fig:GroupSizeExplained}A-D, where only for $N > 3$ a set $\mathbb{D}_i$ of agents disconnected from the focal agent $i$ can exists, i.e. $\mathbb{D}_i = \mathbb{A}\setminus (\mathbb{N}_i \cup \{i\}) \neq \emptyset \text{ for } N > 3$ with $\mathbb{A}$ as the set of all agents of the group.
We confirmed this by computing the average number of neighbors during the simulations (Fig.~\ref{fig:GroupSizeExplained}E).

In summary, for  $N\leq 3$ we have an all-to-all coupling, thus all agents receive the same social input, whereas for $N>3$, centrally located individuals receive "independent" social inputs from different neighbors on different sides, which are not neighbors themselves, i.e. are not directly interacting. 
Thus, for $N > 3$ the centrally located individuals seek a compromise between two independent sources of information. As a consequence, the neighborhood of a focal agent located on the edge and the edge-agent itself, agree less in velocity. This results in slowing down the focal edge-agent, which in turn feeds back on the group behavior. 
To support this explanation we computed the vector product $\mathbf{\Phi}_{\mathbb{N}_i} \cdot  \mathbf{\hat{e}}_{v,i}$ which shows a sudden decrease from $N=3$ to $N=4$ (Fig.~\ref{fig:GroupSizeExplained}F).

\section{Discussion}
We have shown experimental evidence of speed variability in fish and that inertia together with rotational friction explain the reduced turning ability at larger speeds.
With our model that incorporates both, we explored the effect of speed variability on the emergent collective behavior.

Naturally, the decrease of turning rate at higher individual speeds will inhibit individual directional noise, and thus facilitate stronger group polarization. Note, that this effect of speed-dependent turning rate comes on top of previously identified positive impact of higher speeds on group order \cite{Jolles2017}. 
However, not only differences in (average) speed itself, but also differences in individual speed variability for the same average individual speed can result in  differences in polarization between groups.
We find that the local speed correlates strongest with the local polarization at the order-disorder transition, i.e. fixed speed models that investigate this prominent transition miss one of its crucial components.
For example, fixed speed models show at this transition a strong feedback between the maximum susceptibility to perturbations and the self-organized group structure \cite{KlamserRomanczuk2021}, which becomes less pronounced and eventually vanishes for large enough speed variability.
Finally, we unveil a sudden decrease in individual average speed at a threshold group size only present at sufficiently high speed variability, which intrinsically linked to the fundamental structure of the interaction network.

The transition from ordered to disordered motion with speed was reported in experiments \cite{Kent2019, Gautrais2012, Mishra2012}. However, corresponding models incorporating a dependence of turning rate on speed were based on fitting of experimental data and not on the fundamental physics of inertia and rotational friction (see e.g. \cite{Gautrais2012, Mishra2012}).
This order-disorder transition induced by speed might enhance collective computation, as collective gradient sensing reported in golden shiners \cite{Berdahl2013}.
If a model mimics the reported increased/decreased speed for undesired/desired environmental cues (light intensity), a variable speed model that correctly accounts for inertia could enhance the tendency of the collective to stay in the desired environment because there it would be disordered, further decreasing group speed. 

To elaborate the connection to collective computation we stress that the reported maximum of the local correlation between polarization and speed at the disorder-order transition supports the criticality hypothesis \cite{Mora2011, Munoz2018}.
More specifically, it suggests that at the transition, also referred to as the critical point, information encoded in the speed is linked strongest to directional information, i.e. the individuals within the group show the strongest response to directional information via speed adaptations and vise versa.
However, we have also shown that spatial group properties show distinct extrema at the transition for fixed-speed models, which are weakened or can even vanish with increasing speed variability.
This may have important implications regarding observations and conclusions drawn based on investigation of fixed speed models and highlights the complexity of the transition region \cite{KlamserRomanczuk2021}.

The distance regulating force, combining attraction and repulsion is required to obtain a cohesive shoal, and we used here a simple, yet generic form, of this interaction.
For the alignment interaction experimental evidence is species \cite{Katz2011a, Herbert-Read2011, Calovi2018} and group size dependent \cite{Gautrais2012, Zienkiewicz2018, SbragagliaKlamser2020} but the chosen form is as general as possible.
The individual processing of velocity information can be even more elaborate than just taking into account the velocity differences of all neighbors.
An extension was studied by Lemasson et al. \cite{Lemasson2013} where a focal agent only processes information of those neighbors that move significantly faster compared to its total neighborhood.
However, as long as the velocity alignment force acts also on the speed of the agent our qualitative results are expected to hold.

Specific experimental data can be mimicked by a multitude of models which differ strongly in their microscopic interactions \cite{Bastien2020, Romanczuk2009, Couzin2002, SbragagliaKlamser2020}.
However, those models are most often fit to a specific experimental setup, i.e. to a certain group size, tank size and depth, and need to be recalibrated if the setup changes \cite{Gautrais2012}. 
Recently \cite{Jhawar2020} suggested that the experimentally observed decrease in polarization with larger groups is an emergent property of a model with only pairwise interactions.
However, in our model with a low speed relaxation strength $\beta$, we observe the same functional dependence of polarization on group size.
Linking our results again to criticality: only at the disorder-order transition does the polarization saturates for large groups to intermediate values.

The reported sudden speed decrease in our variable speed model at a critical group size of $N=3$, linked to a transition from an all-to-all network to a distributed spatial network,  might offer alternative means to test hypotheses about the underlying interaction network in real animal groups \cite{Ballerini2008a,strandburg2013visual}.
In our model Voronoi-interactions cause the specific size threshold at $N=3$, but for example for k-nearest neighbor interaction 
a group is all-to-all connected up to a threshold size directly set by $k$, i.e. for $N<k$.
However, in order to observe this qualitative change the neighbors have to align their velocity vectors \cite{Herbert-Read2013, Herbert-Read2017}, i.e. also match their speeds instead of only matching their movement direction. There might be also other limitation to this approach, however the emergent speed-structure coupling clearly shows how taking into account variable speed may introduce novel effects at the group level via the self-organized interplay of speed and orientation dynamics and social interactions. Thus, we conclude that extreme caution should be taken when drawing strong conclusions on collective behavior of animal groups based on agent-based models with fixed speed.

\section*{Conflict of Interest Statement}
The authors declare that the research was conducted in the absence of any commercial or financial relationships that could be construed as a potential conflict of interest.

\section*{Author Contributions}
PPK, LGN and PR designed the analysis and wrote the paper. TL and DB provided the experimental data and commented on the draft.

\section*{Funding}
We received financial support by the DFG (German Research Foundation) under BI 1828/2-1, RO 4766/2-1, LA 3534/1-1 and under the Germany’s Excellence Strategy – EXC 2002/1 “Science of Intelligence” – project number 390523135.

\section*{Acknowledgments}
We are thankful to Jens Krause for stimulating discussions.

\section*{Supplemental Data}

\section*{Data Availability Statement}
The experimental data sets are available in previous articles \cite{Bierbach2017, Doran2019, Jolles2020}. The code to run the agent-based model is available at github (\url{https://github.com/PaPeK/swarm-variable-speed}).

%
\bibliographystyle{unsrt}
\bibliography{biblio}
\newpage
\setcounter{page}{1}

\begin{center}
{\LARGE SI Appendix
\\
Supplementary Information
\\
``Impact of Variable Speed on Collective Movement of Animal Groups''
}
\\
\bigskip
Pascal P. Klamser\textsuperscript{1,2}, Luis G\'omez Nava\textsuperscript{1,3}, Tim Landgraf\textsuperscript{3,4}, Jolle W. Jolles,\textsuperscript{5}, David Bierbach,\textsuperscript{3,6,7}, Pawel Romanczuk\textsuperscript{1,2,3,*}
\\
\bigskip
\textbf{1} Institute for Theoretical Biology, Department of Biology, Humboldt Universit\"at zu Berlin, Berlin, Germany. \\
\textbf{2} Bernstein Center for Computational Neuroscience, 10115 Berlin, Germany. \\
\textbf{3} Cluster of Excellence, Science of Intelligence, Technische Universität Berlin, Berlin, Germany,\\
\textbf{4} Department of Mathematics and Computer Science, Freie Universität Berlin, Berlin, Germany. \\
\textbf{5} Center for Ecological Research and Forestry Applications (CREAF), Campus de Bellaterra (UAB), Barcelona, Spain. \\
\textbf{6} Department of Biology and Ecology of Fishes, Leibniz-Institute of Freshwater Ecology and Inland Fisheries, Berlin, Germany. \\
\textbf{7} Faculty of Life Sciences, Albrecht Daniel Thaer-Institute of Agricultural and Horticultural Sciences, Humboldt Universit\"at zu Berlin, Berlin, Germany.
\bigskip
\bigskip
\end{center}
\renewcommand{\theequation}{S\arabic{equation}}
\renewcommand{\thetable}{S\arabic{table}}
\renewcommand\thefigure{S\arabic{figure}}    
\setcounter{equation}{0} 
\setcounter{section}{0} 
\setcounter{figure}{0} 
\renewcommand\thesection{\Roman{section}}    


\section{Details on experimental setups}\label{sec:SI_expSetups}
The experimental data presented in this article was published in Jolles et al. \cite{Jolles2020} (guppies and RoboFish), Bierbach et al. \cite{Bierbach2017} (single mollies) and Doran et al. \cite{Doran2019} (groups of mollies).
In the following we summarize the setups of each study.
\subsection{Individual guppy and RoboFish trajectories}
For this experiments, only adult female individuals (Trinidadian guppies) were used, which have a standard body length of ($31.7 \pm 0.8$ mm).
The single individual trajectories were obtained by putting single fish in a $88$cm $\times$ $88$cm white glass tank.
The behavior was recorded over an observation period of 10 minutes using an acquisition frame rate of 30 FPS.
The videos were then processed using the software BioTracker \cite{Monck2018} to obtain the tracking of each fish.

For the RoboFish trials, a three-dimensional-printed fish replica was used. 
This replica was connected to a two-wheeled robot below the tank (see Fig.~S1 of the SI in \cite{Jolles2020}).
The robot was controlled by a closed-loop system whereby the movements of the fish were identified and fed back to the robot control.
The robot was able to adjust its position and direction of motion to mimic natural responses.
The robot's behavior was based on the zonal model explained in \cite{Couzin2002}.
For more details on the acquisition of the experimental data, see \cite{Jolles2020}.

\subsection{Mollies trajectories: single individual experiments}
The experiments were performed using clonal Amazon mollies using a open field circular tank ($48.5$cm of diameter) made of white plastic.
Single fish were introduced and observed for periods of 5 minutes.
Its behavior and motion were recorded and its positions were acquired with the tracking software Ethovision Version 10.1 (Noldus Information Technologies Inc.).
For more details on the data acquisition, see \cite{Bierbach2017}.

\subsection{Mollies trajectories: group experiments}
For these experiments, 32 fish were used. 
They were assembled in groups of 4 individuals, leading to 8 groups in total.
The fish were adult sized-matched mollies of a standard body lenght of ($6.14 \pm 0.76$ cm).
The experiments were performed in a $60$cm $\times$ $30$cm arena.
The groups were left unperturbed for periods of $5$ minutes, in which their behavior was recorded using an acquisition frame rate of 30 FPS.
The positions of the individuals were obtained from the videos using the tracking software Ethovision XT12 (Noldus Information Technology, Inc.).
For more details on the data acquisition, see \cite{Doran2019}.

\begin{table}[]
  \centering
    \begin{tabular}{l cc cc cc cc cc}
                                            & \multicolumn{2}{c}{RoboFish} & \multicolumn{2}{c}{Guppy} & \multicolumn{2}{c}{Molly} & \multicolumn{2}{c}{Mollies(N=4)} & \multicolumn{2}{c}{Model} \\
                                    \cmidrule[1pt](lr{.7em}){2-3}      \cmidrule[1pt](lr{.75em}){4-5}   \cmidrule[1pt](lr{.75em}){6-7}\cmidrule[1pt](lr{.75em}){8-9}\cmidrule[1pt](lr{.75em}){10-11}
                                                   & $F$ & $\alpha$                 & $F$ & $\alpha$                    & $F$ & $\alpha$               & $F$ & $\alpha$                 & $F$ & $\alpha$\\
    $\frac{d\varphi}{dt}= F/v$                     & 0.1 &                  & 0.1 &                                 &  0.45 &                    & 0.2  &                       & 6.2 &  \\
    $\frac{d\varphi}{dt}= F/(v+\alpha)$            & 4.1 & 0.13               & 10.6 & 0.17                        & 4.74 & 0.16                  & 3.03 & 0.4                  & 29.1 & 0.99 \\
    \end{tabular}
  \caption{\textbf{Statistical model parameters.} Fitting parameters for the fits displayed in Fig.~\ref{fig:expEvidence}.
  }
  \label{tab:modelFitParas}
\end{table}

\begin{table}[]
  \centering
    \begin{tabular}{c l l c c c c c}
           & & & standard & Fig.~\ref{fig:expEvidence} & Fig.~\ref{fig:PhaseTransition} & Fig.~\ref{fig:GroupStructure} & Fig.~\ref{fig:GroupSize}\\
                \cmidrule[1pt](lr{.75em}){4-4}   \cmidrule[1pt](lr{.75em}){5-5} \cmidrule[1pt](lr{.75em}){6-6} \cmidrule[1pt](lr{.75em}){7-7}  \cmidrule[1pt](lr{.75em}){8-8}
    \multirow{5}{*}{\rotatebox[origin=c]{90}{\textbf{single}}}
      & preferred speed & $v_0$ & 1 & & $[0.1, 4]$ & & 2 \\
      & speed relaxation & $\beta$ & 0.2 & & $[0.25, 10.76]$ & $[1, 64]$ & $[0.025, 10.76]$ \\
      & turn friction & $\alpha$ & 1 & 0.1 & & &  \\
      & angular noise & $D_\varphi$ & 1 &  & & &  \\
      & velocity noise & $D_v$ & 0.4 &  & & &  \\
      \hline
    \multirow{5}{*}{\rotatebox[origin=c]{90}{\textbf{collective}}}
      & group size & $N$ & 400 & 1 &   & &  \\
      & alignment strength & $\mu_{alg}$ & 2 & - & & $[0, 5.9]$ &  \\
      & distance strength & $\mu_d$ & 2 & - & & &  \\
      & distance slope & $m_d$ & 2 & - & & &  \\
      & preferred distance & $r_d$ & 1 & - & & &  \\
    \end{tabular}
\caption{Parameters used in simulations.
         The different columns after the "standard" column list parameters which differ from standard parameters for the simulations represented by the respective figures.
  }
  \label{tab:paraSummary}
\end{table}

\section{Derivation of turning dependence on speed}\label{sec:SI_deriveTurning}

Here we show in detail how to derive the dynamics of speed and turning angle from the change in velocity: 
\begin{align}
    \frac{\text{d} \mathbf{v}_i}{\text{d} t} = \frac{1}{m}\mathbf{F}_i
\end{align}
with $\mathbf{v}_i$ as the velocity vector of the object, $m$ its mass and $\mathbf{F}_i$ as the force acting on it.
The velocity vector can be expressed via the speed $v_i$ and the heading angle $\varphi_i$ to $\mathbf{v}_i = v_i [\cos \varphi_i, \sin \varphi_i]^T = v_i\ \hat{\mathbf{e}}_{v, i}$ . 
We can reformulate the velocity dynamics by the speed and heading angle dynamics to 

\begin{align}
\frac{\text{d} \mathbf{v}_i}{\text{d} t}
    =
    \frac{\text{d}}{\text{d} t} (v_i\ \hat{\mathbf{e}}_{v, i})
    &=
    \frac{\text{d} v_i}{\text{d} t} \hat{\mathbf{e}}_{v, i}
    +
    \frac{\text{d} \varphi_i}{\text{d} t} \frac{\text{d} \hat{\mathbf{e}}_{v, i}}{\text{d} \varphi_i} v_i \\
    &=
    \frac{\text{d} v_i}{\text{d} t} \hat{\mathbf{e}}_{v, i}
    +
    \frac{\text{d} \varphi_i}{\text{d} t} \begin{bmatrix}-\sin \varphi_i \\
                                         \cos \varphi_i\end{bmatrix} v_i \\
    &=
    \frac{\text{d} v_i}{\text{d} t} \hat{\mathbf{e}}_{v, i}
    +
    \frac{\text{d} \varphi_i}{\text{d} t}\ \hat{\mathbf{e}}_{\varphi, i}\ v_i\ .
\end{align}
where $\hat{\mathbf{e}}_{v, i} = \big[ \cos{\varphi_i(t), \sin{\varphi_i(t)}} \big]^T$ and $\hat{\mathbf{e}}_{\varphi, i} = \big[ -\sin{\varphi_i(t), \cos{\varphi_i(t)}} \big]^T$ are unitary vectors in the particle's heading and turning directions respectively.
Thus, the change in speed and heading angle is

\begin{align}
    \frac{\text{d} v_i}{\text{d} t} &= \frac{\text{d} \mathbf{v}_i}{\text{d} t} \cdot \hat{\mathbf{e}}_{v, i}
    =\frac{\mathbf{F}_i}{m} \cdot \hat{\mathbf{e}}_{v, i} \\
    \frac{\text{d} \varphi_i}{\text{d} t} &= \frac{1}{v_i} \frac{\text{d} \mathbf{v}_i}{\text{d} t} \cdot \hat{\mathbf{e}}_{\varphi, i}
    =\frac{\mathbf{F}_i}{v_i\ m} \cdot \hat{\mathbf{e}}_{\varphi, i}
\end{align}
where the upper index $T$ indicates transposed vectors.

\section{Neighborhood polarization and speed reduction} \label{sec:SI_neighPolar}

In Order to understand the sudden change in individual speed at group-sizes larger than $N=3$, we show that the dependence of the neighborhood polarization $\Phi_{N_i}$ changes qualitatively at this threshold. For group sizes 

The alignment force acting on a focal agent $i$ from Eq.~\ref{eq:AlignmentForce} is 
\begin{align}
  \mathbf{F}_{i, a}(t)
  = \frac{\mu_{alg}}{|\mathbb{N}_i|} \sum_{j \in \mathbb{N}_i}  \mathbf{v}_{ji}(t) 
  = \frac{\mu_{alg}}{|\mathbb{N}_i|} \sum_{j \in \mathbb{N}_i}  (\mathbf{v}_{j}(t) - \mathbf{v}_{i}(t))\ .
\end{align}
If we now assume that each agent swims with the same speed $v_i(t) = v(t),\ \forall i$, it becomes clear that the alignment force directly depends on the neighborhood polarization $\mathbf{\Phi}_{\mathbb{N}_i}(t) = \frac{1}{|\mathbb{N}_i|} \sum_{j \in \mathbb{N}_i} \frac{\mathbf{v}_j}{|\mathbf{v}_j|} = \frac{1}{|\mathbb{N}_i|} \sum_{j \in \mathbb{N}_i} \mathbf{u}_j$:
\begin{align}
  \mathbf{F}_{i, a}(t)
  &= \frac{\mu_{alg}}{|\mathbb{N}_i|} \sum_{j \in \mathbb{N}_i}  (\mathbf{v}_{j}(t) - \mathbf{v}_{i}(t)) \\
  &= \frac{\mu_{alg} v(t)}{|\mathbb{N}_i|} \sum_{j \in \mathbb{N}_i}  (\mathbf{u}_{j}(t) - \mathbf{u}_{i}(t)) \\ 
  &= \mu_{alg} v(t) \left( \mathbf{\Phi}_{\mathbb{N}_i}(t) - \mathbf{u}_{i}(t)) \right)\ .
\end{align}
The alignment force acts on the speed of the focal agent $i$:
\begin{align}
    \frac{\text{d}v_i(t)}{\text{d}t} &= \beta \big(v_0 - v_i(t)\big) + F_{i, v}(t) + \sqrt{2 D_{v}} \,\, \xi_v(t)\\
     &\propto \mathbf{F}_{i, a}(t) \cdot \mathbf{u}_i(t)
     \propto \left( \mathbf{\Phi}_{\mathbb{N}_i}(t) - \mathbf{u}_{i}(t) \right) \cdot \mathbf{u}_i(t)  \\
     &\propto \mathbf{\Phi}_{\mathbb{N}_i}(t) \cdot \mathbf{u}_i(t) - 1 \\
     &\propto \Phi_{\mathbb{N}_i} \cos \angle_{\mathbf{\Phi}_{\mathbb{N}_i}, \mathbf{u}_i} - 1 \; .
\end{align}

Thus, the stronger the heading direction deviates from the mean neighborhood heading direction $\hat{\mathbf{\Phi}}_{\mathbb{N}_i} = \mathbf{\Phi}_{\mathbb{N}_i} / \Phi_{\mathbb{N}_i}$, the stronger is the speed decreased.
Of course, the neighborhood polarization $\Phi_{\mathbb{N}_i}$ affects the speed, the less polarized the neighborhood the stronger the speed decrease. 

\end{document}